\documentclass{elsart}
\usepackage{amsbsy,amsmath,amssymb}
\begin{document}
\begin{frontmatter}
\title{INSTANTONS FOR DYNAMIC MODELS FROM B TO H}

\author[1]
{Juha Honkonen\corauthref{cor1}}
\ead{Juha.Honkonen@helsinki.fi},
\author[2]{M. V. Komarova}
\ead{komarova@paloma.spbu.ru} and
\author[2]{M. Yu. Nalimov}
\ead{Mikhail.Nalimov@pobox.spbu.ru}
\address[1]{
Theoretical Physics~Division, Department~of~Physical Sciences,
P.O.~Box~64,FIN-00014 University~of~Helsinki, Finland }
\address[2]{
Department of Theoretical Physics, St.~Petersburg University,
Ulyanovskaya 1, St.~Petersburg, Petrodvorets, 198504 Russia }
\corauth[cor1]{corresponding author}

\begin{abstract}
Instanton analysis is applied to models B---H of critical
dynamics. It is shown that the static instanton of the massless
$\phi^{4}$ model determines the large-order asymptotes of the
perturbation expansion of these near-equilibrium dynamic models leading to
factorial growth with the order of perturbation theory.
\end{abstract}

\end{frontmatter}

\maketitle

\section{\label{sec:intro}Introduction}

Large-order asymptotic behaviour of perturbation series of the
paradigmatic static $\phi^4$ field-theoretic model has been
thoroughly explored with the aid of instanton analysis and applied
to the resummation of the asymptotic series brought about by the
standard perturbation expansion~\cite{Lipatov,ZinnBr,TMF}.

However, large-order asymptotics in dynamic field theories
constructed from Langevin equations with the use of the
Martin-Siggia-Rose (MSR) formalism~\cite{MSR} is almost completely
unexplored.

In this paper we use the recently proposed dynamic-instanton
method \cite{Honkonen04} to assess large-order asymptotic
behaviour of dynamic models from B to H in the standard classification
\cite{Hohenberg77}. All these models possess Gibbsian
static limit and thus describe dynamics near thermodynamic equilibrium.
In the majority of these models mode-coupling effects are present requiring significant
development of the method used for model A in Ref. \cite{Honkonen04}

We construct the stationarity equations for the steepest-descent
method and give arguments in favor of existence of the dynamic
instanton solution. Large-order asymptotes of correlation
functions, response functions and critical exponents are shown to
be determined by the instanton solution of the corresponding
static problem, which in all cases considered may be constructed
with the aid of the known static instanton of the massless
$\phi^{4}$ model.

The present article is organized as follows: in Sec.
\ref{sec:dynfieldthO} representation of the MSR field theory in
the functional-integral form for the generic model of critical
dynamics from the corresponding Langevin equation system is
reviewed. Then in Sec. \ref{sec:dinstantonO} we show that dynamic
instanton may be constructed and the steepest-descent evaluation
of the functional integral carried out in the same fashion as for
model A \cite{Honkonen04} with the result that the leading
contribution is given by the corresponding static instanton.
Construction of the dynamic instanton with the aid of the static
instanton of the massless  $\phi^{4}$ model for models B --- H of
critical dynamics is analyzed in Sec. \ref{sec:statinst}.  Results
of this paper are summarized in Sec. \ref{sec:conclusion}.
Calculation of the fluctuation determinant is reviewed in the
Appendix.

\section{\label{sec:dynfieldthO}Field theory for critical dynamics}

Standard models of critical dynamics may be described by the generic Langevin equation
\begin{equation}
\label{LanEq}
\frac{\partial \varphi_a}{\partial t}+\left(\alpha_{ab}+\beta_{ab}\right) \frac{\delta
S}{\delta\varphi_b}=\xi_a\,,
\end{equation}
determining dynamics of fluctuations in a system with the
thermodynamic equilibrium corresponding to the field theory
with the static ''action'' (effective Hamiltonian) $S$.  In
(\ref{LanEq}) the source $\xi$ is a Gaussian random field with
zero mean and the correlation function
\begin{equation}
\label{correlator} \langle\xi_a(t,{\bf x})\xi_b(t',{\bf
x}')\rangle=2\alpha_{ab}\delta(t-t')\delta({\bf x}-{\bf x}')
\end{equation}
chosen to maintain the fluctuation-dissipation theorem. To this end the following conditions are imposed on
the terms of the kinetic coefficient as well:
\begin{equation}
\label{conditions}
\alpha^\top=\alpha\,,\qquad \beta^\top=-\beta\,,\qquad
\frac{\delta \beta_{ab}}{\delta\varphi_a}=0\,.
\end{equation}
The symmetric part of the kinetic coefficient $\alpha$ is assumed to be field-independent and positive-definite.
In equation
(\ref{LanEq}) the static action is assumed to have three-point and four-point interaction terms only and
to contain the action of the massless $\phi^4$ model
\begin{equation}
\label{staticS}
S_4=\frac{1}{2}\,(\nabla\phi)^2+\frac{g}{4!}\,\phi^4\,,
\end{equation}
for which the instanton solution is known \cite{Lipatov}.
Relations (\ref{LanEq}) -- (\ref{conditions}) constitute a compact
form of the description of all standard models of critical
dynamics \cite{Vasiljevnew}. Here and henceforth, all necessary
integrals and sums are implied. In the forthcoming treatment
the full $d$-dimensional Euclidian space ${\bf R}^d$ is assumed for spatial variables,
but for the time variable it is convenient to carry out
most of the analysis on a finite time interval $[t_0,T]$ and pass to the limit
$ t_{0}\to -\infty $ and $ T\to \infty $ at a later stage.

In the MSR approach, functional
integrals for correlation and response functions are calculated
with the ''measure''
\begin{equation}
\label{DDJmeasure}
\mathfrak{M}[\varphi,\varphi']=
\mathcal{D}\varphi \mathcal{D}\varphi ' \sqrt{\det M_0^TM_0}\, e^{-\overline{S}}\,,
\end{equation}
where the dynamic action for the model (\ref{LanEq})-- (\ref{conditions})
may be written symbolically  \cite{MSR} as
\begin{equation}
\label{DDJaction}
\overline{S}=-\varphi'_a \alpha_{ab}\varphi'_b
+\varphi'_a\left[\frac{\partial\varphi_a}{\partial t}
+\left(\alpha_{ab}+\beta_{ab}\right)\frac{\delta S}{\delta\varphi_b}\right]\,.
\end{equation}
We stick to the choice of Ref. \cite{Honkonen04} of the determinant factor in representation
(\ref{DDJmeasure}) in the field-independent form with the self-adjoint operator
$M_0^TM_0$, where the operator $M_0$ is generated by the free-field part of dynamic action
(\ref{DDJaction}) as
\begin{equation}
\label{M0}
\left(M_{0}\right)_{ab}=\delta_{ab}\frac{\partial }{\partial t}
+\left(\alpha_{ac}+\beta_{ac}^{(0)}\right)K_{cb}\,.
\end{equation}
Here, $\beta^{(0)}$ is the field-independent part of the
skew-symmetric term of the kinetic coefficient, whereas $K_{ab}$ is the kernel of the
quadratic form $S_0$ of the static action:
\begin{equation}
\label{S0}
S_0={1\over 2}\,\varphi_aK_{ab}\varphi_b\,,
\end{equation}
and is thus, by definition, symmetric and positive.
The choice of the determinant factor in measure (\ref{DDJmeasure})
implies that the interaction functional $\overline{S}_I$
is taken in the normal form, which in the perturbation
theory is equivalent to the condition that there are no graphs with closed
loops of the retarded propagators $\left(M_0^{-1}\right)_{ab}(t,{\bf x};t',{\bf x}')$ attached
to the interaction vertex. This amounts to vanishing retarded free-field propagator at coinciding times,
and no field-dependent factors to the determinant $\det M_0^TM_0$ are generated.

We remind that initially in the ordinary perturbation theory the boundary conditions on the time
axis fix $ \varphi'(T)=0 $, but leave $ \varphi (T) $, $ \varphi '(t_{0})$ and $ \varphi '(T)$
arbitrary. In the instanton approach the
space of integration $E(\Delta)$ in the functional integral is chosen using the properties of the
quadratic form of the expansion of dynamic action at the saddle point and therefore it is in
general different from that of the ordinary perturbation theory, the construction being similar to that for model A
\cite{Honkonen04}.

\section{\label{sec:dinstantonO}Instanton analysis for the generic model of critical dynamics}

We are dealing with a dynamic model which has several
coupling constants which, however, are proportional to some powers of
a single parameter, e.g. $\epsilon $, at the fixed point of the
renormalization group equation). Therefore, it seems most reasonable to
base the large-order analysis on the number of loops rather
than powers of different coupling constants.  This may be
effected by extracting a common factor $u$ from all coupling
constants and then counting large orders by powers of $u$.
Thus, we cast the interaction part of the static action in the
form
\begin{equation}
\label{g_scaling}
v_{abc}\varphi_a\varphi_b\varphi_c+g_{abcd}\varphi_a\varphi_b\varphi_c\varphi_d=
u\overline{v}_{abc}\varphi_a\varphi_b\varphi_c+u^2\overline{g}_{abcd}\varphi_a\varphi_b\varphi_c\varphi_d\,.
\end{equation}
The task then amounts to a steepest-descent calculation
of the parametric integral
\begin{equation}
\label{Norder0}
\frac{1}{2\pi i}\oint \frac{du}{u} \frac{\displaystyle\iint \!\mathcal{D}\varphi
\mathcal{D}\varphi '
\Phi_{a_1}(t_1, {\bf{x_1}})\ldots
\Phi_{a_k}(t_k,{\bf{x_k}})\, e^{\displaystyle-\overline{S}-N
\ln u}}{\displaystyle\iint\!  \mathcal{D}\varphi
\mathcal{D}\varphi '\, 
e^{\displaystyle-\overline{S}_0}}\,,
\end{equation}
where $\Phi =\{\varphi,\,\varphi '\}$, expressing the $N$th order
contribution to perturbation expansion in $u$ of the $k$-point
Green function. In representation (\ref{Norder0}) the normalizing
determinant has been written in the form of a functional integral
with the free-field part of the dynamic action in order to make
explicit cancellation of Jacobians related to various changes of
variables. As shown in Ref. \cite{Honkonen04}, an instanton
solution for expression (\ref{Norder0}) for model A may be
constructed in close relation to the instanton solution of the
corresponding equilibrium model. In this Section we show that a
similar procedure allows to reduce the large-order analysis of the
dynamic functional integral (\ref{Norder0}) to that of the static
problem determined by the static action $S$.

It is convenient to extract the large factor $N$ in the exponential of the numerator of expression
(\ref{Norder0}) by a suitable change of variables accompanied by the same transformation in the denominator to avoid
uninteresting Jacobi determinants: $\varphi_a\rightarrow
\sqrt{N}{\varphi}_a$, $\varphi'_a\rightarrow
\sqrt{N}{\varphi}'_a$, $u\rightarrow u/\sqrt{N}$. To simplify the forthcoming
calculation of the fluctuation integral around the saddle
point we also scale out the dependence on the variable $u$ by
$\varphi_a\rightarrow
{\varphi}_a/(iu)$, $\varphi'_a\rightarrow {\varphi}'_a/(iu)$.
As a result of these transformations we arrive --- up to the
factor $N^{(N+k)/2}(iu)^{-k}$ --- at the representation
\begin{equation}
\label{Norder}
\frac{1}{2\pi i}\oint \frac{du}{u}
\frac{\displaystyle\iint \!\mathcal{D}\varphi \mathcal{D}\varphi '
\Phi_{a_1}(t_1, {\bf{x_1}})\ldots
\Phi_{a_k}(t_k,{\bf{x_k}})\,
e^{\displaystyle-N\left(-\overline{S}/u^2+ \ln
u\right)}}{\displaystyle\iint\!  \mathcal{D}\varphi
\mathcal{D}\varphi '\, 
e^{\displaystyle N\overline{S}_0/u^2}}
\end{equation}
where the dynamic action $\overline{S}$ (as well as the static
action $S$) are independent of the loop-counting variable $u$
and these can be obtained from the initial ones by
change of the coupling constants $ g_{abcd} \to -\overline
g_{abcd}$, $ v_{abc}\to -i\overline v_{abc} $.

The stationarity equations of the method of steepest descent
assume the form
\begin{align}
\label{statEqPhi} \frac{\delta\overline{S}}{\delta\varphi_a}&=
-{\partial \varphi'_a\over\partial
t}+\varphi'_b\,\left(\alpha_{bc}+\beta_{bc}\right) \frac{\delta^2
S}{\delta \varphi_c\delta \varphi_a}+\varphi'_b\,
\frac{\delta\beta_{bc}}{\delta\varphi_a}\frac{\delta S}{\delta \varphi_c}=0,\\
\label{statEqPhiPrime}
\frac{\delta\overline{S}}{\delta\varphi'_{a}}&=
-2\alpha_{ab}\varphi'_b+{\partial
\varphi_a\over \partial t}+ \left(\alpha_{ab}+\beta_{ab}\right)\frac{\delta S}{\delta\varphi_b}=0\,,\\
\label{statEqG} \overline{S}&=-{u^2\over 2}\,.
\end{align}
As in static theory, equations (\ref{statEqPhi}) --
(\ref{statEqG}) have a nontrivial solution with negative
$u^{2}$ only. The dynamic instanton $\varphi_{D}$ for the
basic field is the nontrivial solution of the equation
\begin{equation}
\label{stat1.2}
-{\partial \varphi_a\over \partial t}+
\left(\alpha_{ab}-\beta_{ab}\right)\frac{\delta
S}{\delta\varphi_b}=0.
\end{equation}
Substitution of this solution to equation (\ref{statEqPhiPrime})
leads to the nontrivial instanton $\varphi'_{D}$ for the
auxiliary field, whose equation we quote in the two equivalent
forms used later:
\begin{equation}
\label{dPrime} \varphi'_{D\,a}=\frac{\delta S}{\delta\varphi_a}
=\left(\alpha^{-1}\right)_{ab}\left({\partial\varphi_b\over\partial
t}+ \beta_{bc}\frac{\delta S}{\delta\varphi_c}\right)\,.
\end{equation}
With the use of the instanton equation (\ref{stat1.2}) and the
antisymmetry property of the kinetic coefficient $\beta_{ab}$ it
may be readily seen that the stationarity equation
(\ref{statEqPhi}) is fulfilled on solution (\ref{dPrime}).

In addition to equations (\ref{statEqPhi}), (\ref{statEqPhiPrime})
and (\ref{statEqG}) variation on the boundaries of the time
interval produces [assuming fixed initial field $\varphi(t_0,{\bf
x})$] the condition
\begin{equation}
\label{statEqb} \varphi '(T,{\bf x})=0\,,
\end{equation}
which -- due to the first equation (\ref{dPrime}) -- singles
out the solution approaching the static instanton at the final time instant.
Indeed, according to relations (\ref{dPrime}) and (\ref{statEqb}) the
dynamic instanton $\varphi_{D}$ is a function conforming to the
Cauchy condition for the {\em final} time instant
$\varphi_{D}(T,{\bf x})=\varphi_{st}({\bf x})$, where
$\varphi_{st}$ is the static instanton solution for the static
action $S$. It should be noted that the iterative solution of the
instanton equation (\ref{stat1.2}) also has the property
$\lim\limits_{t\to-\infty}\varphi_{D}(t,{\bf x})=0$, because
--- due to the sign changes in comparison with
(\ref{statEqPhiPrime}) --- the Green's function of
linear part of equation (\ref{stat1.2}) vanishes infinitely
far in the past.  A similar equation for the model A was
investigated in \cite{Honkonen04}, where it was shown  that
the solution of this equation on the infinite time range for
negative $u^{2}$ tends to zero at $t\to -\infty$ and to
$\varphi_{st}$ at $t\to \infty$.

Substitution of the solution $\varphi_D$ and $\varphi'_D$ in
dynamic action (\ref{DDJaction}) reveals that the dynamic
action on the dynamic instanton solution gives rise to a time
integral of a total time derivative (the expression $\partial
S/\partial t=(\partial \varphi /\partial t)(\delta S/\delta
\varphi )$ was used) asymptotically coincides with the static
action on the static instanton:  $$
\overline{S}\left(\varphi_D,\varphi'_D\right)=S\left(\varphi_{st}\right)-S\left(\varphi_0\right)
\xrightarrow[t_0 \to
-\infty]{\hbox{}}S\left(\varphi_{st}\right)\,, $$
here $\varphi_{0}({\bf x})=\varphi_{D}(t_0,{\bf x})$.
Equation (\ref{stat1.2}) gives rise to iterative solution with an advanced Green's function,
therefore the field $\varphi_0$ vanishes in the limit $t_0\to -\infty$

Similar the
third saddle-point equation (\ref{statEqG}) on solution
(\ref{stat1.2}) and (\ref{dPrime}) reduces to the
following  equation for the limiting values of the dynamic
instanton
\[
S\left(\varphi_{st}\right)-
S\left(\varphi_0\right)=-\frac{u^2}{2}\,,
\]
and thus recovers the stationarity equation for the static
instanton
\begin{equation}
\label{staticG}
S\left(\varphi_{st}\right)=-\frac{u^2}{2}\,,\qquad t_0\to -\infty\,.
\end{equation}
Thus, as for model A, we arrive at the conclusion that the exponential factor in our steepest descent
analysis of Green function (\ref{Norder}) as well as
pre-exponential factor asymptotically are the same as in the
corresponding equilibrium static theory, provided there is a
nontrivial solution of equation (\ref{stat1.2}). Such a
solution may be constructed in the same fashion as in model A
as the usual tree-graph solution of the nonlinear equation
(\ref{stat1.2}) with the given final Cauchy condition
$\varphi_{D}(T,{\bf x})=\varphi_{st}({\bf x})$ instead
of an initial condition.

In the integration space
of the functional integral of the $k$-point Green functions
(\ref{Norder})
the degeneracies of the instanton due to the usual
dilatation invariance and
translation invariance in coordinate space
may be removed
with the aid of the same unit decomposition~\cite{Lipatov,Zinn,Suslov}
as in the static instanton theory with the corresponding conditions imposed on
on the final value of the integration field $\varphi(T,{\bf x})$.

Thus, we arrive at the Green function
\begin{equation}
\label{NorderI}
\frac{1}{2\pi i}\oint \frac{du}{u}
\frac{\displaystyle\iint \!\mathcal{D}\varphi \mathcal{D}\varphi '
\Phi_{a_1}(t_1, {\bf{x_1}})\ldots
\Phi_{a_k}(t_k,{\bf{x_k}})\,{\rm I}\,
e^{\displaystyle-N\left(-\overline{S}/u^2+ \ln
u\right)}}{\displaystyle\iint\!  \mathcal{D}\varphi
\mathcal{D}\varphi '\,
e^{\displaystyle N\overline{S}_0/u^2}}
\end{equation}
where ${\rm I}$ stands for all
contributions from the degeneracy-lifting unit decomposition.
At the leading order in $N$ we replace all $\Phi
(t_i,{\bf{x}_i})$ by $\{\varphi_{D}, \varphi'_{D}\}$ in the
pre-exponential factor. In the stationary limit
$t_0\to-\infty$ the initial value of the instanton
$\varphi_{D}(t_0) $ vanishes and in the exponential the static
instanton action is recovered at leading order in $N$.

The most subtle point of the saddle-point approach, i.e. the calculation of the fluctuation
determinant at the instanton solution, is outlined in the Appendix. The main conclusion is
-- as in the case of the model A \cite{Honkonen04} -- that
the fluctuation determinant coincides with that of the static theory.
The analysis of large-order contributions to
correlation and response functions also proceeds in the same way
as for model A and we do not dwell on it here.

Thus, we conclude that in the translation-invariant in time case the asymptotic properties of
the dynamic model at leading order in $N$ are determined by the static instanton solution which
leads to factorial growth of the large-order contributions as in the static instanton analysis.
Therefore, the large-order behaviour of an arbitrary quantity $F$ (correlation or response function or
critical index) may be expressed as
\begin{equation}
\label{c1}
F^{[N]}= C\,N!a_{M}^{N}N^{b}\,,
\end{equation}
where $ F^{[N]} $ is the $N\,$th order contribution to $F$ of the expansion in the parameter $ e $
($ e $ is the coupling constant $ g $ or the dimensional regularization parameter $ \epsilon $). In
(\ref{c1}) the constant $a_{M}$ depends on the model only ($M=A,B,C,...$), whereas the exponent $ b
$ and the amplitude factor $ C $ -- either a constant or a function of coordinate and time
arguments -- depend on the quantity $F$ as well. See, for instance, the analysis of the factor $
C $ for a generic function in the static theory presented in Ref. \cite{TMF1}).

As in the case of the model A \cite{Honkonen04}, when expression (\ref{c1}) is used for contributions to
correlation or response functions, the $ \Phi $ fields in the pre-exponential factor of
(\ref{NorderI}) have to be replaced by the static instanton $\varphi_{st}$ for the fields $\varphi $ and put
equal to zero for the fields $\varphi '$ at the leading order in $1/N$. The
dynamic properties of the Green functions are thus determined by the first correction in $1/N$ \cite{Honkonen04}.

\section{\label{sec:statinst}Static and dynamic instantons for
models B --- H}

We have shown in Sec. \ref{sec:dinstantonO} that at the leading order the results of the
dynamic-instanton calculation are
determined by the static instanton corresponding to the static action $S$. In this Section we review the results
of the static-instanton analysis, which in all cases are based on the long-known static instanton of the
massless $\phi^4$ model.

\subsection{\label{sec:statinst:B}Model B}

In model B only one field is present, the static action is that of the
massless $\phi^4$ model
(\ref{staticS})  and the symmetric kinetic coefficient is
proportional to the Laplace operator:
\begin{equation}
\label{alphaB}
\alpha=-\lambda \nabla^2\,.
\end{equation}
Since the static action is exactly the same as in model A and the large-order behaviour of the perturbation
expansion is determined by the static instanton, the
conclusions about this behaviour are the same as for model A:
in model B expression (\ref{c1}) is valid  with
the constant $ a_{B}= a_{0}$  known from
the analysis of the static theory \cite{Lipatov,Zinn}
[for the action (\ref{staticS}) $a_{0}=1/(16\pi ^{2}) $].

Although the explicit form of the dynamic instanton was
not used in the analysis presented above it would be useful for the
investigation of the amplitudes $ C $ in relation (\ref{c1}) for the
dynamic Green functions. Therefore, we remind how the dynamic instanton may be
constructed. A convergent iterative solution of the equation
(\ref{stat1.2}) for model B can be written in analogy with
model A \cite{Honkonen04}, viz.
\begin{equation}
\nonumber
\phi_{D}^{(0)}(t,{\bf x})=\int d{\bf x}'\widetilde G(T-t,{\bf
x}-{\bf x}')\phi_{st}({\bf x}')
\end{equation}
can be taken as the zeroth-order approximation. The next orders
of the expansion are constructed as the usual tree-graph solution
of equation (\ref{stat1.2}) with the aid of the advanced
kernel of the linear operator $ (\partial _{t}-\lambda \nabla
^{4}) $, which can be represented in the form
\begin{equation}
\label{tG} \widetilde G(t,{\bf x})=-\Theta (-t)\int\limits _{-\infty
}^{\infty }dy\frac 1{(2\pi )^{d}\sqrt {4\pi \lambda t}}\left(\frac
{\pi }{iy}\right)^{d/2}e^{-\,\frac {\displaystyle y^{2}}{\displaystyle 4\lambda t}\,+\frac {\displaystyle
i{\bf
x}^{2}}{\displaystyle 4y}}.
\end{equation}

\subsection{\label{sec:statinst:C}Model C}

In model C there are two components of the basic field: let $\varphi_1=\phi$ and $\varphi_2=m$, and the
static action is of the form:
\begin{equation}
\label{staticC}
S_{C}={1\over 2}\,(\nabla\phi)^2+{g\over 4!}\phi^4+{m^2\over
2}+ {1\over 2}\,v_2m\phi^2\,.
\end{equation}
The kinetic coefficient is still symmetric and diagonal
\begin{equation}
\label{alphaC}
\alpha=
\left(
\begin{array}{cc}
\Gamma& 0\\
0& -\lambda \nabla^2
\end{array}
\right)\,.
\end{equation}
To keep track of the number of loops we put $ g=u^{2} $ and $
v_{2}=\varkappa u $. Then the static instanton is determined
by the set of equations
\begin{align}
\label{statC1}
-\nabla^2\phi-{1\over 6}\phi^3-i\varkappa m\phi&=0\,,\\
\label{statC2}
m-{i\over 2}\varkappa \phi^2&=0\,,\\
\label{statC3} {S_{C}}{\renewcommand{\arraystretch}{.7}\Bigl|
\begin{array}{c}\\ {\tiny g=-1}\\{\tiny v_2=-i\varkappa} \end{array}}=-{u^2\over 2}\,,
\end{align}
and thus may be constructed with the use of the usual instanton solution $\phi_{st}$ of the
massless $\phi^4$ model:
\begin{align}
\label{4instanton}
-\nabla^2\phi_{st}-{1\over 6}\,\phi_{st}^3&=0\,,\\
{1\over 2}\int\!d{\bf x}\left(\nabla\phi_{st}\right)^2\,\phi_{st}-{\tilde g\over 4!}
\,\int\!d{\bf x}\,\phi_{st}^4&=-\frac {u^2}2\,,
\end{align}
with $\tilde g=3\varkappa ^2-1$.
To use large-order asymptotes for
investigation of the $ \epsilon $ expansion we write
$ \varkappa =\varkappa (\epsilon )=v_{2*}/\sqrt {g_{*}} $ (
where $ v_{2*} $ and $ g_{*} $ are the coordinates of a fixed
point of the renormalization-group flow equations).

The stability of the fixed point in the model considered depends
on the value of the critical index of the specific heat $\alpha $.
If the index $ \alpha <0$, then $ v_{2*} =0$ at the
stable fixed point.  In this case $ \tilde g=-1 $ and the
constant $ a=a_{C}$ in expression (\ref{c1}) for this model is
the same as in the model A and the usual static $ \phi ^{4} $
model \cite{Lipatov,ZinnBr}:  $ a_{C}=a_{0}$.

In the opposite case, when index $ \alpha >0 $ we have
$g_{*}\sim \epsilon $ and $v_{2*}^{2}\sim \epsilon $ at the stable
fixed point giving rise to a nontrivial $ \epsilon $ expansion for
the parameter $\varkappa=v_2/u$ \cite{Vasiljevnew}:
\begin{equation}
\label{kappaeksp}
\varkappa =\varkappa ^{(0)}+\varkappa ^{(1)}\epsilon
+\varkappa ^{(2)}\epsilon ^{2}+...
\end{equation}
Therefore, the constant $ a_{C} $ in relation (\ref{c1}) is now
determined by the equation
\begin{equation}
\nonumber \frac 1{a_{0}}=\frac {\left[1-3\varkappa (\epsilon )^{2}\right]}{a_{C}}.
\end{equation}
Taking into account that in the $\epsilon$ expansion
asymptotic structure (\ref{c1}) appears as
\begin{equation}
\label{error}
F^{[N]}\epsilon ^{N}=N!Ca_{C}^{N}N^{b}\epsilon ^{N}
\end{equation}
we see that the only contribution to the constant $ a_{C} $
comes from the
$\varkappa ^{(0)}$ term of expansion (\ref{kappaeksp}). In this case, therefore,
\begin{equation}
\nonumber
a_{C}={\frac {a_{0}}{1-3\varkappa ^{(0)2}}}\,.
\end{equation}
The coefficient $\varkappa ^{(1)}$  from expression (\ref{kappaeksp})
determines the correction to the amplitude factor $ C $ of the form
\begin{equation}
\nonumber
\exp \left[\frac {6\varkappa ^{(0)}\varkappa
^{(1)}}{1-3\varkappa ^{(0)2}}\right]\,,
\end{equation}
whereas $\varkappa ^{(2)}$ and higher-order terms of the $ \epsilon $
expansion
of $\varkappa (\epsilon)$ are irrelevant for the leading order in $ 1/N $.

Let us note that actually we are considering here a $ \sqrt {\epsilon} $
expansion, since the loop-counting parameter
$ u\sim \sqrt {\epsilon}$ at the leading order,
and due to the scaling $ \phi ,m\to \phi /(iu),
m/(iu) $ the $u$ expansion of
correlation functions with an odd number of $ m$ fields contains
odd powers of $ u $ whereas those with an even number of $ m$
fields contain even
powers of $ u $ only. Nevertheless, we may use the usual form
(\ref{error}) of the perturbation expansion taking into account simple additional
normalization in the functional-integral representation
(\ref{Norder}) for the Green functions.

The fluctuation determinant is also the same as in the
massless static $\phi^4$ model, because it is given by
the Gaussian integral with the quadratic form
\begin{equation}
\label{d2SC} {1\over 2}\,\delta^2S={1\over
2}\,\delta\phi\left(-\nabla^2+{g\over
2}\,\phi_{st}^2+v_2m_{st}\right)\delta\phi +\delta\phi
v_2\phi_{st}\delta m+{1\over 2}\,\delta m^2\,.
\end{equation}
Here, the gaussian integral over $\delta m$ gives the
determinant of the unit operator and produces the Gaussian
integral over $\delta\phi$ with the quadratic form
\begin{equation}
\label{d2SCphi}
{1\over 2}\,\delta^2S={1\over 2}\,\delta\phi\left(-\nabla^2+{g-3v_2^2\over 2}\,\phi_{st}^2\right)\delta\phi
={1\over 2}\,\delta\phi\left(-\nabla^2-{\tilde
gu^{2}_{st}\over 2}\,\phi_{st}^2\right)\delta\phi
\end{equation}
which is exactly that of the massless $\phi^4$
model. A subsequent integration over $ \delta u $ yields the
additional  factor $ 1/\sqrt {1-3\varkappa ^{2}} $.

Equation (\ref{stat1.2}) for the dynamic instanton here assume the form
\begin{align}
\label{dynC}
\frac{\partial \phi}{\partial t}
&=\Gamma\left(-\nabla^2\phi-{1\over 6}\phi^3-i\varkappa m\phi
\right)\,,\\
\nonumber
\frac{\partial m}{\partial t}&=-\lambda
\nabla^{2}\left(m-{i\over 2}\varkappa \phi^2 \right)\,.
\end{align} The zeroth-order approximation of the iterative
solution can be written as
\begin{align}
\nonumber
\phi _{D}^{(0)}(t,{\bf x})&=\int d{\bf x}'\,G(T-t,{\bf x}-{\bf
x}')\phi _{st}({\bf x}'), \\
\nonumber
m _{D}^{(0)}(t,{\bf x})&=\frac {i\varkappa }2\int d{\bf
x}'\,G(T-t,{\bf x}-{\bf x}')\phi _{st}^{2}({\bf x}'),
\end{align}
where
\begin{equation}
\label{G}
G(t,{\bf x})=\Theta (-t)\frac 1{\left(4\pi \Gamma |t|\right)^{d/2}}\exp
\left(-\frac {\displaystyle {\bf x}^{2}}{\displaystyle 4 \Gamma |t|}\right)
\end{equation}
is the advanced diffusion kernel which determines the propagator in
the tree-graph solution here.

\subsection{\label{sec:statinst:D}Model D}

Model D has the same static action (\ref{staticC}) as model C,
but the kinetic coefficient is slightly different:
\begin{equation}
\nonumber
\alpha=
\left(
\begin{array}{cc}
-\lambda \nabla^2& 0\\
0& -\lambda _{1} \nabla^2
\end{array}
\right)\,.
\end{equation}
The matrix element $\alpha_{11}$ of this kinetic coefficient is the
same as in model B.

Therefore, the analysis of the previous section \ref{sec:statinst:C} is valid here in full extent
with the only replacement of the diffusion propagator $ G $ for the $\phi $ field from (\ref{G}) by
the "hyperdiffusion" propagator $\widetilde G $ from relation (\ref{tG}) in the first equation of
set (\ref{dynC}) and subsequently in the tree-graph expansion for $ \phi _{d} $. Therefore, the constant
$a_D$ is equal to $a_c$.

\subsection{\label{sec:statinst:F}Model F}

In model F there are two complex conjugate fields: $\varphi_1=\psi$,
$\varphi_2=\psi ^{*}$ and a real field $\varphi_3=m$. The
static action is of the form:
\begin{equation}
\label{staticF}
S_{F}=\vert\nabla\psi\vert^2+{g\over
6}\vert\psi\vert^4+{m^2\over 2}+ v_2m\vert\psi\vert^2 \,.
\end{equation}
The kinetic coefficient $ (\alpha +\beta )$ is determined by the matrices
\begin{equation}
\label{alphaF}
\alpha= \left(
\begin{array}{ccc}
0&\lambda _{\psi}& 0\\
\lambda _{\psi}&0&0\\
0&0& -\lambda _{m}\nabla^2
\end{array}
\right)\,,
\end{equation}
\begin{equation}
\label{betaF}
\beta= \left(
\begin{array}{ccc}
0&iv_{3}&iv_{4}\psi \\
-iv_{3}&0&-iv_{4}\psi ^{*}\\
-iv_{4}\psi ^{*}&iv_{4}\psi & 0
\end{array}
\right)\,.
\end{equation}

The static instanton is determined by the set of equations
(\ref{statC1}), (\ref{statC2}) and (\ref{statC3}) of the model C,
where the filed $\phi $ is treated as a two-component one with the
components  $ \psi ^{*}/\sqrt {2} $ and $ \psi /\sqrt {2}$ and $
\phi ^{2}=\psi ^{*}\psi /2$. Then the constant $a= a_{F} $ in relation
(\ref{c1}) for this model was determined in section
\ref{sec:statinst:C}.  The fluctuation determinant is also the
same as in the model C and coincides with that of the
massless static $\phi^4$ model.

Equation (\ref{stat1.2}) for the dynamic instanton is here
\begin{align}
\label{dynF}
\frac{\partial \psi}{\partial t} &=(\lambda _{\psi
}+iv_{3})\left(-\nabla^2\psi+{g\over
3}\vert\psi\vert^2\psi+v_2m\psi \right)-iv_{4}\psi
\left(m+v_{2}\vert\psi\vert^2\right)\,,
\nonumber \\
\nonumber
\frac{\partial \psi ^{*}}{\partial t}&=(\lambda _{\psi
}+iv_{3}) \left(-\nabla^2\psi ^{*}+{g\over
3}\vert\psi\vert^2\psi ^{*}+v_2m\psi ^{*}\right)+iv_{4}\psi ^{*}
\left(m+v_{2}\vert\psi\vert^2\right)\,,\\
\frac{\partial m}{\partial t}&=-\lambda _{m}
\nabla^{2}\left(m+v_2\vert\psi\vert^2\right)-iv_{4}\nabla(\psi
^{*}\nabla \psi -\psi \nabla \psi ^{*})\,
\end{align}
with coefficients $g=-1$, $ v_{2}=-i\varkappa $, $v_{4}\to
-iv_{4}/u$. The iterative solution can be written here in
analogy with the model C using the advanced propagator $ G $
from (\ref{G}).

\subsection{\label{sec:statinst:E}Model E}

Model E is F model with $ v_{2}=v_{3}=0 $. Thus, it has been
analyzed in the previous section. The constant $a_{E} $ in relation
(\ref{c1}) is equal
to $ a_{0} $ due to $ v_{2}=0 $. The static instanton for the field $ m $
is $ m_{st}=0 $, but a nontrivial dynamic
instanton for this field may be obtained from equations (\ref{dynF}) due to
mode coupling.

\subsection{\label{sec:statinst:G}Model G}

In model G there are two real vector fields: let
$\varphi_a=\phi_{a}$ and
$\varphi_{3+a}=m_{a}$, where $a=1,2,3$. The static action is of the form:
\begin{equation}
\label{staticG}
S_{G}={1\over 2}\,(\nabla\phi)^2+{g\over 4!}\,\phi^4+{m^2\over
2}\,,
\end{equation}
and the kinetic coefficient is determined by the relations
\begin{equation}
\label{alphaG}
\alpha= \left(
\begin{array}{cc}
\lambda _{\phi}& 0\\
0& -\lambda _{m}\nabla^2
\end{array}
\right)\,.
\end{equation}
\begin{equation}
\label{betaG}
\beta_{ab}=0\,,\quad
\beta _{a\,{3+b}}=v_{2}\epsilon _{abc}\phi _{c},\quad
\beta _{3+a\,3+b}=v_{2}\epsilon _{abc}m_{c}\,,
\end{equation}
where $\epsilon _{abc}$ is the completely antisymmetric tensor
and all the indices $a$, $b$, $c$ assume values 1,2,3.

The static instanton solution for this model is $ \phi =\phi _{st} $, $ m=0 $.
Therefore $ \tilde g=-1 $ and the constant $a_{G}$ in relation
(\ref{c1}) coincides with the static one for the massless
$\phi^4$ model: $ a_{G}=a_{0} $. The
fluctuation determinant is also the same as in the
$\phi^4$ model.

The dynamic-instanton equation (\ref{stat1.2}) for this model can be solved as
described in the preceding sections.

\subsection{\label{sec:statinst:H}Model H}

The static action of H model is of the form:
\begin{equation}
\label{staticH}
S_{H}={1\over 2}\,(\nabla\phi)^2+{g\over 4!}\phi^4+{m^2\over
2}+ {1\over 2}\,v_2m\phi^2+{c\over 2}{\bf v}^{2}\,,
\end{equation}
with the scalar fields $ \phi $ and  $ m $ and the transversal vector field $ {\bf v}$.
The static instanton solution for the scalar fields turns out to be the the same as in model C,
whereas $ {\bf v}_{st}=0 $. Therefore, the fluctuation determinant has been also
discussed in section \ref{sec:statinst:C}. But in fact the critical
dynamics is described by the simplified H$ _{0} $ model without the field $ m $,
whose static action is
\begin{equation}
\label{staticH0}
S_{H_{0}}={1\over 2}\,(\nabla\phi)^2+{g\over 4!}\phi^4+{c\over
2}{\bf v}^{2}\,.  \end{equation} This means that the constant
$ a_{H} $  here as well as the
fluctuation determinant are the same as in the usual $ \phi ^{4} $
model: $ a_{H}=a_{0} $.

The kinetic coefficient is given by the matrices
\begin{equation}
\label{alphaH0}
\alpha= \left(
\begin{array}{cc}
-\lambda _{\phi }\nabla^2& 0\\ 0& -\lambda _{v}\nabla^2
\end{array} \right)\,,\qquad
\beta= \left(
\begin{array}{cc}
0& v_{2}\overrightarrow \nabla \phi\\ -v_{2}\overleftarrow
\nabla \phi& 0 \end{array} \right)\,,
\end{equation}
where the transverse projection operator for the vector field $ {\bf v} $ is implied.

Equation (\ref{stat1.2}) for the dynamic instanton in the H$_0$ model assumes the form
\begin{align}
\nonumber
\frac{\partial\phi}{\partial t} &=-\lambda _{\phi }\nabla
^{2}\left(-\nabla^2\phi-{1\over 6}\phi^3\right)-i\frac
{v_2c}u{\bf v\nabla }\phi \,,\\
\nonumber
\frac{\partial {\bf v}}{\partial t}&=-\lambda c\nabla^{2}{\bf
v}+i\frac {v_2}u\phi \nabla ^3\phi \,.
\end{align}
The iterative solution may be constructed with the use of the
advanced propagators (\ref{tG}) and (\ref{G}) [the propagator
(\ref{G}) for the vector field $ {\bf v} $ taken in a
convolution with the transversal projection operator,
however].

\section{\label{sec:conclusion}Conclusion}

In this paper we have shown that the instanton method is applicable
to large-order asymptotic analysis for all the near-equilibrium
standard models B -- H of critical dynamics in a form which is a
generalization of the recently proposed dynamic-instanton approach in
model A \cite{Honkonen04}. The factorial growth of the large-order coefficients of
the perturbation expansion
\begin{equation}
\label{c2}
F^{[N]}= N!\,Ca_{M}^{N}N^{b}
\end{equation}
has been proved for any quantity $F$ calculable in the form of a
perturbative series. The constants $ a_{M} $ in expression
(\ref{c2}) has been determined for all models mentioned. The
exponent $ b $ in relation (\ref{c2}), however, depends on the
quantity $F$ considered. For any particular $F$ it may be readily
calculated with the use of the results of the present
analysis.

For instance, it may readily be proved that the exponent $ b $
in the $\varepsilon $-expansion contribution to the dynamic index $z$ is
determined by the large-order asymptote of the expansion of the fixed-point
value $ g_{*} $ and the second term of the perturbation series of
the two-point correlation function. Therefore, in the $ O(n)$
symmetric dynamic theory the relation
\begin{equation}
\nonumber
b=3+\frac n2
\end{equation}
follows for the dynamic index $z$.

The fluctuation contribution to the amplitude factor $ C $ in relation (\ref{c2})
has been analyzed for all models B -- H as well.

The work was supported in part by the Nordic Grant for Network
Cooperation with the Baltic Countries and Northwest Russia No.
FIN-20/2003, and by the Academy of Finland (Grant No. 207939).
M.V.K. and M.Yu.N. acknowledge the Department of Physical Sciences
of the University of Helsinki for kind hospitality. The authors
are grateful to A.N. Vasil'ev for fruitful discussions.

\section*{\label{sec:appendix}Appendix}
Let us outline the calculation of the fluctuation determinant at
the instanton solution along the lines proposed
for the model A \cite{Honkonen04}.  Since we have extracted the $u$ dependence
from the dynamic action as a prefactor in Eq. (\ref{NorderI}), due
to the stationarity conditions (\ref{statEqPhi}) --
(\ref{statEqG}) the contribution from the fluctuations of $u$  is
independent of the field fluctuations and factorized as well. The
remaining nontrivial fluctuation integral over $\delta\varphi_a$
and $\delta\varphi'_a$ may be written as
\begin{multline}
\delta \Sigma= \Biggl(\displaystyle\iint
\!\mathcal{D}\delta\varphi\mathcal{D}\delta\varphi' \exp\Biggl\{
\delta\varphi'_a \alpha_{ab}\delta\varphi'_b
-\delta\varphi'_a\left[\delta_{ac}
\frac{\displaystyle\partial}{\displaystyle \partial t} +
\left(\alpha_{ab}+\beta^{(0)}_{ab}\right)\right]\delta\varphi_c\Biggr\}
\Biggr)^{-1}\\
\times
\displaystyle\iint\!\mathcal{D}\delta\varphi\mathcal{D}\delta\varphi'\,
{\rm I} \exp
\Biggl\{\delta\varphi'_a \alpha_{ab}\delta\varphi'_b\\
-\delta\varphi'_a\left[\delta_{ac}
\frac{\displaystyle\partial}{\displaystyle \partial t} +
\left(\alpha_{ab}+\beta_{ab}\right) \frac{\displaystyle\delta^2
S}{\displaystyle\delta\varphi_b\delta\varphi_c}
+\frac{\displaystyle\delta
\beta_{ab}}{\displaystyle\delta\varphi_c}
\frac{\displaystyle\delta S}{\displaystyle\delta\varphi_b}\right]
\bigg |_{\varphi_{D}}\!\!\! \delta\varphi_c\\
-\frac{\displaystyle 1}{\displaystyle 2}\,
\delta\varphi_c\varphi'_{D\,a}\biggl[
\left(\alpha_{ab}+\beta_{ab}\right) \frac{\displaystyle\delta^3
S}{\displaystyle\delta\varphi_b\delta\varphi_c\delta\varphi_d}\\
+2\frac{\displaystyle\delta
\beta_{ab}}{\displaystyle\delta\varphi_c}
\frac{\displaystyle\delta^2
S}{\displaystyle\delta\varphi_b\delta\varphi_d}
+\frac{\displaystyle\delta^2
\beta_{ab}}{\displaystyle\delta\varphi_c\delta\varphi_d}
\frac{\displaystyle\delta S}{\displaystyle\delta\varphi_b}\biggr]
\bigg|_{\varphi_{D}}\!\!\! \!\delta\varphi_d \Biggl\}\,,
\label{variation}
\end{multline}
where $\beta^{(0)}_{ab}$ is the field-independent part of the
kinetic coefficient $\beta$ (we remind that the kinetic
coefficient $\alpha$ is assumed to be field-independent). The
change of variables
\begin{equation}
\label{varchange} \delta\varphi'_a=\delta \psi'_a+
\left(\alpha^{-1}\right)_{ab}\left({\partial\delta\varphi_b\over\partial
t}+ \beta_{bc}\frac{\delta^2
S}{\delta\varphi_c\delta\varphi_d}\,\delta\varphi_d
+\frac{\displaystyle\delta
\beta_{bc}}{\displaystyle\delta\varphi_d}
\frac{\displaystyle\delta
S}{\displaystyle\delta\varphi_c}\,\delta\varphi_d \right)
\end{equation}
together with the antisymmetry of the kinetic coefficient $\beta$,
the instanton equation (\ref{stat1.2}) and the second
equality in relation (\ref{dPrime})
then allows to express the quadratic form of the
exponential of the numerator of fluctuation integral
(\ref{variation}) as
\begin{multline}
\label{d2S} -{1\over 2}\,\delta^2_{\{ \varphi,\varphi'\}
}\overline{S}=
\delta \psi'_a\alpha_{ab}\delta \psi'_b\\
-\delta \psi'_a\left\{-{\partial\delta\varphi_a\over\partial t}+
\left[
\left(\alpha_{ab}-\beta_{ab}\right)\frac{\displaystyle\delta^2
S}{\displaystyle\delta\varphi_b\delta\varphi_c}
-\frac{\displaystyle\delta
\beta_{ab}}{\displaystyle\delta\varphi_c}
\frac{\displaystyle\delta S}{\displaystyle\delta\varphi_b}
\right]\bigg |_{\varphi_{D}}\!\!\! \delta\varphi_c\right\}\\
-\frac{\displaystyle 1}{\displaystyle 2}\, \delta\varphi_{st\,a}
\frac{\displaystyle\delta^2
S}{\displaystyle\delta\varphi_a\delta\varphi_b}\bigg
|_{\varphi_{st}}\!\!\! \delta\varphi_{st\,b} +\frac{\displaystyle
1}{\displaystyle 2}\, \delta\varphi_{0\,a}
\frac{\displaystyle\delta^2
S}{\displaystyle\delta\varphi_a\delta\varphi_b} \bigg
|_{\varphi_0}\!\!\! \delta\varphi_{0\,b}\,,
\end{multline}
where $\delta\varphi_{st\,a}=\delta\varphi_a\vert_{t=T}$ and
$\delta\varphi_{0\,a}=\delta\varphi_a\vert_{t=t_0}$.  To calculate
the fluctuation integral perturbatively, we write --- using the
change of variables (\ref{varchange}) --- the normalization factor
in terms of the perturbation expansion implied in (\ref{d2S}):
\begin{multline}
\label{d2S0} -{1\over 2}\,\delta^2_{\{ \varphi,\varphi'\}
}\overline{S}_0= \delta \psi'_a\alpha_{ab}\delta \psi'_b -\delta
\psi'_a\left[-\delta_{ac}{\partial\over\partial t}+
\left(\alpha_{ab}-\beta^{(0)}_{ab}\right)K_{bc}\right]\delta\varphi_c\\
-\frac{\displaystyle 1}{\displaystyle 2}\,
\delta\varphi_{st\,a}K_{ab}\delta\varphi_{st\,b}
+\frac{\displaystyle 1}{\displaystyle 2}\, \delta\varphi_{0\,a}
K_{ab}\delta\varphi_{0\,b}
\end{multline}
with the same free-field part for the perturbative calculation as
in (\ref{d2S}).
This representation gives rise to the propagators $\Delta_{12}=\Delta^\top_{21}$
and the correlation function $\Delta_{11}$
determined by the operator $M_0$ [defined by equation (\ref{M0})], its transpose
$M_0^\top$ and the correlation function of the Langevin source (\ref{correlator}) as
\begin{align}
\label{Bpropagators}
\left(\Delta_{12}\right)_{ab}(t,{\bf x};t',{\bf x}')&=\left(M_0^{-1}\right)_{ab}(t,{\bf x};t',{\bf x}')\,,\\
\left(\Delta_{11}\right)_{ab}(t,{\bf x};t',{\bf x}')&=
2\left(\Delta_{12}\alpha\Delta_{21}\right)_{ab}(t,{\bf x};t',{\bf x}')\,,
\end{align}
where $\Delta_{12}$ is the retarded Green's function of the operator $M_0$
(and, correspondingly, $\Delta_{21}$ the advanced Green's function).

Inspection of the Fourier transform of the propagator
$\left(\Delta_{12}\right)_{ab}(t,{\bf x};t',{\bf x}')$ shows that
that in the time-space representation it is rapidly decaying  at $ t-t'\to\infty$,
which, together with the
initial condition $\lim\limits_{t-t'\to
0_{+}}\left(\Delta_{12}\right)_{ab}(t,{\bf x};t',{\bf
x}')=\delta_{ab}\delta({\bf x}-{\bf x}')$, is basically all
what is needed for an analysis similar to that for model A
\cite{Honkonen04}.

Let us introduce for the numerator of relation (\ref{variation})
the following decomposition of the integration variable
\begin{align}
\label{intvar}
\delta\varphi_a&=\overline{\delta\varphi}_a+\delta\varphi^T_a,
\end{align}
where $\overline{\delta\varphi}$ consists of the standard Gaussian integration variable
constructed with the aid of the {\em advanced} Green's function
$\left(\Delta_{21}\right)_{ab}$ and
the solution $h$ of the homogenous equation $\left(M_0^\top\right)_{ab} h_b(t,{\bf x})=0$
such that $\overline{\delta\varphi}_a(T,{\bf x})=0$, i.e.
\begin{equation}
\label{var1}
\overline{\delta\varphi}_a=
\left(\Delta_{11}\right)_{ab}\eta_b+\left(\Delta_{21}\right)_{ab}\eta'_b+h_a\,,
\end{equation}
where $\eta_a$ and $\eta'_a$ are fields vanishing at the
boundaries (i.e. at $\vert{\bf x}\vert\to\infty$, $t=T$ and
$t=t_0$). This choice is accompanied by the prescription
\begin{equation}
\label{var2}
\delta\psi'_a=\left(\Delta_{12}\right)_{ab}\eta_b
\end{equation}
for the auxiliary field.

The second term $\delta\varphi^T$ in the decomposition (\ref{intvar})
is chosen as the solution of the homogeneous equation
\begin{equation}
\label{Tequation}
\left[-\delta_{ac} {\partial\over\partial t} +
\left(\alpha_{ab}-\beta_{ab}\right)\frac{\displaystyle\delta^2
S}{\displaystyle\delta\varphi_b\delta\varphi_c}
-\frac{\displaystyle\delta
\beta_{ab}}{\displaystyle\delta\varphi_c}
\frac{\displaystyle\delta S}{\displaystyle\delta\varphi_b}
\right]\bigg |_{\varphi_{D}}\!\!\! \delta\varphi^T_c(t,{\bf x})=0
\end{equation}
with the final value
$\delta\varphi^T_a(T,{\bf x})=\delta\varphi_{st\,a}({\bf x})$
to conform to the boundary condition.
The solution of
equation (\ref{Tequation}) can be constructed perturbatively using the tree-graph
expansion of $\varphi_{D}$.

The integration space for the normalization factor in the denominator of relation (\ref{variation})
is constructed with the aid of a decomposition similar to (\ref{intvar}) with
the same $\overline{\delta\varphi}$ and $\delta\psi'_a$. In this case, however, the second term
$\delta\varphi^T$  in the decomposition may be expressed explicitly as
the {\em spatial} convolution of
the advanced Green's function
$\left(\Delta_{21}\right)_{ab}(t,{\bf x};T,{\bf x}')$ (note the
fixed time argument) and the function $\delta\varphi_{st\,a}({\bf x})$
and thus is the solution of the free-field counterpart of equation (\ref{Tequation}):
$$
\left[-\delta_{ac} {\partial\over\partial t}
+\left(\alpha_{ab}-\beta^{(0)}_{ab}\right)K_{bc}\right]\delta\varphi^T_c(t,{\bf
x})=0\,.
$$
With the use of the asymptotic properties of the integration variables constructed in this manner
it may be shown that in the limit
$T\to \infty$ the final (static) fluctuations factorize completely
(as in the case of model A):
\begin{multline}
\label{variationT}
\delta \Sigma= \left[\displaystyle\int
\!\mathcal{D}\delta\varphi_{st} \exp\left(-\frac{\displaystyle
1}{\displaystyle 2}\delta\varphi_{st\,a}
K_{ab}\delta\varphi_{st\,b}\right)
\right]^{-1}\\
\times \displaystyle\int \!\mathcal{D}\delta\varphi_{st}\,{\rm
I}\, \exp\left(-\frac{\displaystyle 1}{\displaystyle
2}\delta\varphi_{st\,a}\frac{\displaystyle\delta^2 S}
{\displaystyle\delta\varphi_a\delta\varphi_b}\bigg
|_{\varphi_{st}}\!\!\! \delta\varphi_{st\,b}\right)
\Biggl(\displaystyle\iint
\!\mathcal{D}\overline{\delta\varphi}\mathcal{D}\delta \psi'
\exp\Biggl\{\delta \psi'_a \alpha_{ab}\delta \psi'_b\\
-\delta \psi'_a\left[-
\frac{\displaystyle\partial\overline{\delta\varphi}_a}{\displaystyle
\partial t} + \left(\alpha_{ab}-\beta^{(0)}_{ab}\right)K_{bc}
\overline{\delta\varphi}_c\right]+\frac{\displaystyle
1}{\displaystyle 2}\, \overline{\delta\varphi}_{0\,a}
K_{ab}\overline{\delta\varphi}_{0\,b}\Biggr\}
\Biggr)^{-1}\\
\times\displaystyle\iint\!\mathcal{D}\overline{\delta\varphi}\mathcal{D}
\delta \psi'\,  \exp \Biggl\{\delta \psi'_a \alpha_{ab}\delta
\psi'_b -\delta \psi'_a\left[-
\frac{\displaystyle\partial\overline{\delta\varphi}_a}{\displaystyle
\partial t} +
\left(\alpha_{ab}-\beta_{ab}\right)\frac{\displaystyle\delta^2
S}{\displaystyle\delta\varphi_b\delta\varphi_c}\right.\\
\left.
-\frac{\displaystyle\delta
\beta_{ab}}{\displaystyle\delta\varphi_c}
\frac{\displaystyle\delta S}{\displaystyle\delta\varphi_b}
\right]\bigg |_{\varphi_{D}}\!\!\! \overline{\delta\varphi}_c
+\frac{\displaystyle 1}{\displaystyle
2}\,\overline{\delta\varphi}_{0\,a} \frac{\displaystyle\delta^2
S}{\displaystyle\delta\varphi_a\delta\varphi_b}\bigg
|_{\varphi_{0}}\!\!\! \overline{\delta\varphi}_{0\,b} \Biggr\}\,,
\end{multline}
where
$\overline{\delta\varphi}_{0\,a}=\overline{\delta\varphi}_a\vert_{t=t_0}$.
On the right-hand-side of (\ref{variationT}) the first two factors
yield the static fluctuation determinant including the
degeneracy-lifting unit decomposition. A non-trivial dependence on
the initial time instant $t_0$ remains in (\ref{variationT})
through the non-zero initial value of the fluctuation field
$\overline{\delta\varphi}_{0}$. When the translation invariance
with respect to time is restored by passing to the limit $t_0\to
-\infty$, however, the initial value of the fluctuation field
vanishes: $\overline{\delta\varphi}_{0}\to 0$ due to the structure
of the integration field (\ref{intvar}) and the attenuation
property of the advanced Green function. In a manner similar to that
used for model A \cite{Honkonen04}
it may then be shown perturbatively that in this limit the dynamic
contribution to the fluctuation integral (\ref{variationT}) tends
to unity and thus we arrive at the fluctuation integral
\begin{equation}
\label{variationS} \lim\limits_{T\to\infty\atop t_0\to
-\infty}\delta \Sigma= {\displaystyle\int
\!\mathcal{D}\delta\varphi_{st}\,{\rm I}\,
\exp\left(-\frac{\displaystyle 1}{\displaystyle
2}\,\delta\varphi_{st\,a} \frac{\displaystyle\delta^2
S}{\displaystyle\delta\varphi_a\delta\varphi_b}\bigg
|_{\varphi_{st}}\!\!\! \delta\varphi_{st\,b}\right) \over
\displaystyle\int \!\mathcal{D}\delta\varphi_{st}
\exp\left[-\frac{\displaystyle 1}{\displaystyle
2}\,\delta\varphi_{st\,a} K_{ab}\delta\varphi_{st\,b}\right]}
\end{equation}
completely determined by the static theory.

\end{document}